\documentclass[prl,
twocolumn,superscriptaddress,
preprintnumbers,
amssymb, amsmath, floatfix]{revtex4-1}
\usepackage{graphicx}

\newcommand{\nn}{\ensuremath{\nonumber}}

\renewcommand{\vec}[1]{\ensuremath{\boldsymbol{\mathrm{#1}}}}

\def\independenT#1#2{\mathrel{\setbox0\hbox{$#1#2$}%
\copy0\kern-\wd0\mkern4mu\box0}}

\begin{document}
\title{Resonating valence bonds and Fermi surface reconstruction: The resistivity in the underdoped cuprates}%
\author{Phillip E. C. Ashby}
\email{ashbype@mcmaster.ca}
\affiliation{Department of Physics and Astronomy, McMaster University, Hamilton, Ontario, Canada L8S 4M1}

\author{J. P. Carbotte}
\email{carbotte@mcmaster.ca}
\affiliation{Department of Physics and Astronomy, McMaster University, Hamilton, Ontario, Canada L8S 4M1}
\affiliation{The Canadian Institute for Advanced Research, Toronto, Ontario, Canada M5G 1Z8}

\begin{abstract}
The pseudogap phase of the underdoped cuprates is the host to a variety of novel electronic phenomenon.  An example is the dc-resistivity which shows metallic behaviour in the ab-plane, while the c-axis response is insulating.  We apply a model, originally formulated by Yang, Rice, and Zhang, to study the resistivity in the pseudogap phase. This model is able to reproduce the qualitative features of the resistivity, including the systematic deviations from linear behaviour for the in-plane conductivity, and the insulating behaviour along the c-axis.  We compare this to the predictions of the arc model and find similar qualitative behaviour.  We find that the most important element in understanding the resistivity is the reconstruction of the Fermi surface, which puts strong restrictions on the number of quasiparticles allowed to participate in dc-transport.
\end{abstract}

\maketitle

The psuedogap is the precursor phase from which high-$T_c$ superconductivity condenses in the underdoped cuprates.  Understanding the pseudogap phase is believed to be central for deducing the mechanism that underpins high-$T_c$ superconductivity \cite{Timusk:1999fk}. To study the low lying excitations in this phase, it is essential to understand the character of its Fermi surface (FS). Photoemission (ARPES) in the psuedogap phase observes segments of FS along the Brillouin Zone diagonals.  These segments were originally interpreted as disconnected segments of FS and became known as Fermi arcs \cite{Kanigel:2007bh}. Recently, they have been resolved as pockets with small spectral weight along the aniferromagnetic Brillouin zone (AFBZ) boundary \cite{Yang:2011uq}.  Evidence for pockets is compounded by the observation of quantum oscillations, which requires a closed FS \cite{Doiron-Leyraud:2007qf}.  The nature of the FS reconstruction in the pseudogap phase has been the subject of much debate and has lead to a number of theoretical proposals.

One class of theories of the pseudogap phase involve competing order parameters whose fluctuations serve to reconstruct the FS into the observed pockets.  These models involve broken translational symmetry, such as charge or spin density waves. Other theories of the pseudogap begin from the idea of preformed Cooper pairs that lack off-diagonal long range order \cite{Norman:2007vn}.  Indeed, superconducting fluctuations have been seen to exist above $T_c$ in the form a disordered vortex liquid \cite{Xu:2000ys}. However, these fluctuations fail to persist to the temperature scale associated with pseudogap behaviour \cite{Li:2010zr}. An alternative approach is that the pseudogap arises naturally as one dopes a Mott insulator, which then leads to both insulating and strongly correlated electronic behaviour \cite{Lee:2006kx}.  This was the approach that Yang, Rice, and Zhang (YRZ) adopted when they put forward their model of the underdoped cuprates in 2006 \cite{Yang:2006ly}.

The YRZ model consists of an ansatz for the coherent piece of the single particle Greens function in the pseudogap phase.  It is based on results for Anderson's resonating valence bond (RVB) spin liquid \cite{ANDERSON:1987fk,Zhang:1988uq}.  In this description, as the Mott insulating state is approached a gap opens on the AFBZ boundary.  This gap is an RVB spin gap and appears as an energy scale separate from superconductivity.

Since its debut, the YRZ model has been shown to capture many properties of the cuprates that were considered anomalous. In particular, it has been shown to give good agreement with Raman Spectra \cite{Valenzuela:2007kx, LeBlanc:2010vn}, ARPES \cite{Yang:2009ys,Yang:2011uq}, specific heat \cite{LeBlanc:2009zr}, penetration depth \cite{Carbotte:2010ly}, and tunnelling spectroscopy \cite{Yang:2010ve}.  Perhaps most remarkably is that the YRZ model captures all of this behaviour with doping as the only free parameter.  

\begin{figure}
  \includegraphics[width=0.9\linewidth]{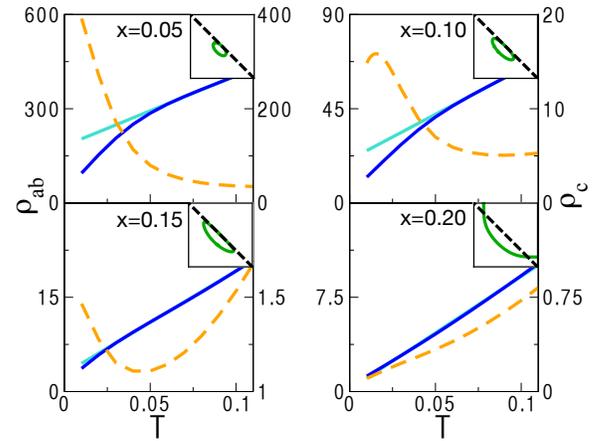}
  \caption{(Color online) In plane (solid blue, in units $e^{-2}d$) and c-axis (dashed orange, in units $e^{-2}t_\perp^{-2}d^{-1}$) resistivities as a function of temperature in the YRZ model for dopings x = 0.05, 0.10, 0.15 and 0.2. The solid turquoise lines are fits to the linear T region.  The c-axis resistivity shows insulating like behaviour, and the in-plane resistivity shows a systematic deviation from linearity induced by the pseudogap energy scale.  The insets show the Fermi surfaces  (shown for one quarter of the BZ, AFBZ boundary indicated by the dashed line) given by the YRZ Greens function.}
  \label{fig:1}	
\end{figure}

A related model, the arc model, has also been shown to capture the qualitatively correct behaviour in the underdoped cuprates (see for example Refs \cite{LeBlanc:2010vn, LeBlanc:2009zr,Carbotte:2010ly}).  In this model, a gap is placed on the antinodal portion of the metallic fermi surface, leaving Fermi arc segments.  The length of the remaining FS serves the same role doping does in the YRZ model.  Thus, it seems that the qualitative behaviour of many of these experiments  can be captured by any model that gives the disappearance of FS along the antinodal direction. It is crucial to understand in what circumstances the YRZ and arc models  provide qualitatively different pictures of the electronic response in the pseudogap phase.

In this letter, we compute the in-plane (ab), and out-of-plane (c-axis) resistivity using both the YRZ model and the arc model for the pseudogap phase of the underdoped cuprates.  This calculation is of an essentially different character than the previous studies, as it requires us to add ingredients not present in the initial formulation of YRZ.  Specifically, the resistivity relies on the scattering rate, and the extension to the c-axis needs some assumption to be made about the interlayer tunnelling matrix element.  After adding these ingredients we show that the YRZ and arc models agree qualitatively with dc-transport experiments. Although these two models have different underlying physics, they give qualitatively similar predictions for the resistivity in the underdoped cuprates.

{\emph{YRZ Model ---}} The YRZ ansatz for the coherent part of the electron Greens function for a doping $x$ is
\begin{align}
\label{eq1}G(\vec{k},\omega) = \frac{g_t}{\omega-\xi({\vec{k}})-\Delta_{PG}^2/[\omega+\xi^0(\vec{k})]},
\end{align}
where $g_t = 2x/(1+x)$ is a Gutzwiller renormalization factor, $\xi(\vec{k}) = \xi^0(\vec{k})-4t'(x)\cos(k_x)\cos(k_y)-2t''(x)(\cos(2k_x)+\cos(2k_y))-\mu_p$ is the tight binding dispersion on a square lattice out to third nearest neighbours, and $\xi^0(\vec{k}) = -2t(x)(\cos(k_x)+\cos(k_y))$ is the first nearest neighbour term.  $\mu_p$ is a chemical potential determined from the Luttinger sum rule.  The RVB gap, $\Delta_{PG}$, plays the role of the pseudogap and has d-wave symmetry, i.e., $\Delta_{PG} = \Delta_0(1-x/x_c)(\cos(k_x)-\cos(k_y))$.  We use values for all parameters in the YRZ model as they appear in Ref. \cite{Yang:2006ly}.  We work in units where $\hbar = k_\textrm{B} = 1$, and all energies are measured in units of $t_0$ (see Ref. \cite{Yang:2006ly}). 

Equation (\ref{eq1}) would be the Greens function for a superconductor if $\xi^0(\vec{k}) = \xi(\vec{k})$.  This difference causes the gap to open on the surface defined by $\xi^0(\vec{k})$, which in our case is the AFBZ.  This gap naturally reconstructs the large FS into pockets and can give rise to interesting electronic response.  Given this Greens function we can compute the conductivity, and hence, the resistivity.

The dc-conductivity is given by the Kubo formula
\begin{align}
\sigma_{ij} = -\lim_{\omega\rightarrow0}\left[\frac{\textrm{Im}(\Pi_{ij}(\omega))}{\omega}\right].
\end{align}
The current-current correlation function, $\Pi_{ij}(\omega)$, is evaluated at the one-loop level
\begin{align}
\label{eq2}\Pi_{ij}(\omega)  = e^2T\sum_{\vec{k},\nu_n} \left[v_i(\vec{k}) G(\vec{k},\nu_n) v_j(\vec{k}) G(\vec{k},\omega+\nu_n)\right],
\end{align}
where $v_i$ is the $i^\textrm{th}$ component of the velocity and the $\nu_n$ are Matsubara frequencies.  For the ab-conductivity we use $v_x = d\xi(\vec{k})/dk_x$.  For the c-axis conductivity, we replace $v^2$ with $t_\perp^2(\vec{k})d^2$, where $d$ is the interlayer distance and $t_\perp(\vec{k}) = t_\perp\left(\cos(k_x)-\cos(k_y)\right)^2$ is the interlayer tunnelling matrix element.  This matrix element was first suggested by Anderson \cite{Chakravarty:1993qf}. It was used in the nodal liquid model of the cuprates, which reproduced the insulating-like behaviour associated with the pseudogap \cite{Levchenko:2010bh}.  The final component that we need to compute the conductivity is the scattering rate, $\Gamma$, which broadens the spectral densities.  To make contact with experiments we follow the suggestion of reference \cite{Ito:1993dq} and take a temperature dependent scattering rate.  For our calculations we include a small residual scattering and take $\Gamma = 0.01 + 2\pi\lambda T$, with $\lambda = 0.3$.  These choices leave us with doping as our only tuneable parameter.

Figure \ref{fig:1} shows the in-plane and out-of-plane resistivity as a function of temperature for a few values of doping.  The qualitative agreement with experiments is good for both the in-plane and out-of-plane results (see references \cite{Ito:1993dq,Takenaka:1994cr}).  At high temperatures the in-plane resistivity is linear. As the pseudogap opens (moving down in doping), there is a progressively larger deviation from the linear behaviour.  This deviation from linearity is a loss of metallically associated with the shrinking of the FS.  The c-axis resistivity shows strong insulating behaviour at low values of doping, and becomes increasingly metallic as optimal doping is approached, just as in the experiments.

\begin{figure}
  \includegraphics[width=0.8\linewidth]{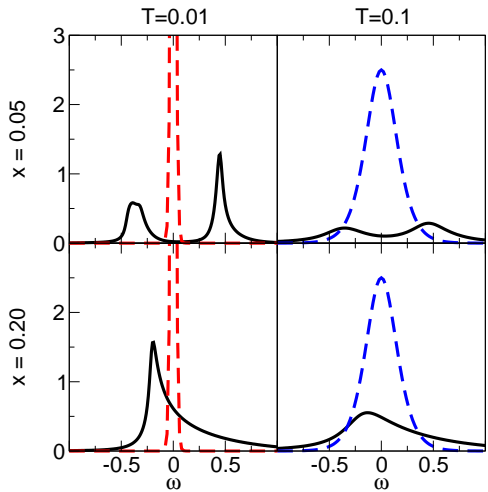}
  \caption{(Color online) Joint density of states involved in the conductivity calculation for $x =0.05$ (highly underdoped) and $x=0.20$ (optimal doping) at low and high temperatures. The red dashed curves are the thermal factors for the low temperature case and peak height $\approx$ 25.   The blue dashed curves are the the thermal factors for the high $T$ case.  As the JDOS becomes more flat (psedogap fingerprint blurred out) we return to a more metallic-like behaviour.}
  \label{fig:2}	
\end{figure} 

To understand these results, we found it instructive to introduce a joint density of states (JDOS).  The JDOS can be thought of as a reference function, and its computation is the same as Eq. (\ref{eq2}), but with the velocity operators omitted.  This function is shown for the optimally doped and underdoped cases in figure \ref{fig:2} for the c-axis.  The only difference in the JDOS between the ab- and c-axis conductivity is the factor $\left(\cos(k_x)-\cos(k_y)\right)^2$ from the tunnelling matrix element.  This factor suppresses the spectral weight near zero frequency.

At optimal doping the JDOS contains only a single peak, and the dependence on the resistivity is controlled solely by thermal factors.  At low temperatures, when the thermal factors are more sharply peaked, we get a higher conductivity and hence a lower resistivity.  As we increase the temperature the thermal factors broaden, we get a smaller conductivity (larger resistivity).  The underdoped case is more interesting. In the underdoped case the opening of the pseudogap causes the peak to split into two pieces, separated by an energy on the order of the gap scale.  The thermal factors behave as before, but there is no spectral weight remaining in the low $T$ case.  This gapping of the region near $\omega = 0$ naturally gives insulating behaviour.   As the temperature is increased the JDOS broadens due to increased scattering, and our resistivity falls back down accordingly.  Within the YRZ model the resistivity is determined by this interplay between thermal factors and the structure of the density of states coming from the pseudogap.

\begin{figure}
  \includegraphics[width=0.8\linewidth]{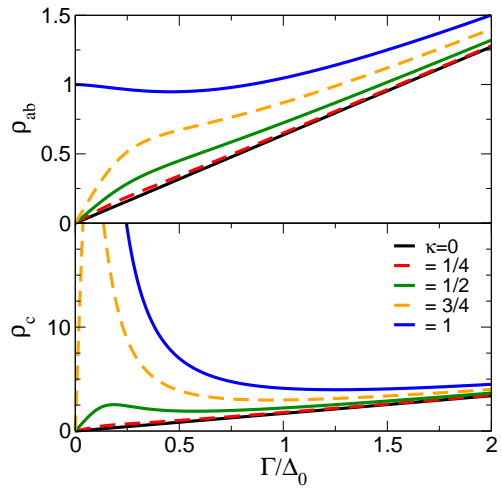}
  \caption{(Color online) The in-plane resistivity (units of $\frac{d\pi\Delta_0}{e^2Nv_F^2}$, $N$ is the density of states at the Fermi energy, and $v_F$ is the Fermi velocity.) and c-axis resistivity (units of $\frac{\pi\Delta_0}{e^2Ndt_\perp^2}$) in the arc model for dimensionless pairing strengths $\kappa$ = 0, 0.25, 0.5, 0.75, and 1.}
  \label{fig:3}	
\end{figure}

{\emph{Arc Model ---}}
We now turn our attention to the arc model.  In the arc model, we start with a large FS as in the optimally doped case.  By hand we place a gap, $\Delta$, on part of the FS starting from the antinodal direction.  The amount gapped out is paramaterized by an angle, $\theta_c$, which plays the role of doping.  There are then two contributions to the conductivity: a `free electron' part ($\sigma_{\textrm{arc}}$) from the remaining FS, and an `interband' part ($\sigma_\Delta$) from the gapped region of the FS.  We then compute the conductivity for a scattering rate $\Gamma$ for both the in-plane, and c-axis.  We obtain (units as in Figure 3) $\sigma^{(ab)} = \sigma^{(ab)}_{\textrm{arc}}+\sigma^{(ab)}_{\Delta}$ and $\sigma^{(c)} = \sigma^{(c)}_{\textrm{arc}}+\sigma^{(c)}_{\Delta}$ with
\begin{align}
\sigma^{(ab)}_{\textrm{arc}} = \frac{2}{\Gamma}\left(\frac{\pi}{4}-\theta_c\right),
\end{align}
\begin{align}
\sigma^{(ab)}_{\Delta} = \frac{\lambda}{\Delta}\left(E\left(2\theta_c\Big|\lambda^2\right)-\frac{\lambda^2}{2}\frac{\sin(4\theta_c)}{\sqrt{1-\lambda^2\sin^2(2\theta_c)}}\right),
\end{align}
\begin{align}
\sigma^{(c)}_{\textrm{arc}} =  \frac{1}{\Gamma}\left[\frac{3}{4}\left(\frac{\pi}{4}-\theta_c\right)-\frac{1}{4}\sin(4\theta_c)-\frac{1}{32}\sin(8\theta_c)\right],
\end{align}
\begin{align}
\nn\sigma^{(c)}_{\Delta} = \frac{\lambda \Gamma^4}{\Delta^5}&\left[\left(2+\frac{\Delta^2}{\Gamma^2}\right)E\left(2\theta_c\Big|\lambda^2\right)-2F\left(2\theta_c\Big|\lambda^2\right)\right.\\
&\left.-\frac{\lambda^2}{2}\frac{\sin(4\theta_c)}{\sqrt{1-\lambda^2\sin^2(2\theta_c)}}\right],
\end{align}
where $\lambda = \Delta/\sqrt{\Delta^2+\Gamma^2}$, $F(x|\lambda^2)$ is the incomplete elliptic integral of the first kind, and $E(x|\lambda^2)$ is the incomplete elliptic integral of the second kind.

Unlike in YRZ, in the arc model the arc length is independent of the magnitude of the pseduogap.  To make contact with the experiments and YRZ we chose to trade $\theta_c$ for a new dimensionless variable $\kappa = \Delta/\Delta_0$, where $\Delta_0$ is the magnitude of the pesudogap at 0 doping.   In principle the function $\theta_c(\kappa)$ is a strictly increasing function of $\kappa$ which could be obtained by fitting the arc model to experimental data, if one desired. To highlight the essential features, and for simplicity, we chose $\theta_c = \kappa \pi/4$.  The resulting curves are shown in figure \ref{fig:3}.  Since the scattering rate $\Gamma \propto T$, the x-axis may be read as temperature.  

\begin{figure}
  \includegraphics[width=0.8\linewidth]{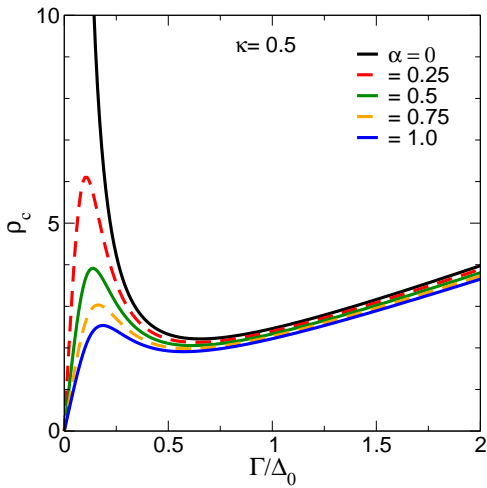}
  \caption{(Color online) The c-axis resistivity with the weight given by the remaining FS as a tuneable parameter, $\alpha$.  Specifically, $\alpha$ enters as $\alpha\sigma^{(c)}_{\textrm{Arc}}+\sigma^{(c)}_\Delta$.  This figure shows how the FS contributes to the c-axis conductivity.  The FS still gives an important contribution, even though the tunnelling  matrix element to gaps out the nodal excitations.}
  \label{fig:4}	
\end{figure}

Although the precise details of the curves differ from both the YRZ model and the experiments, the broad features are still captured by this simple model. The in-plane resistance is linear at high $T$, and shows systematically larger deviations as the paring strength is increased.  The case $\kappa = 1$ is distinct from the other cases, and corresponds to the universal limit first discussed by Lee \cite{Lee:1993nx}.  The c-axis resistivity shows progressively insulating behaviour as we proceed further into the  underdoped regime. This is due to the loss of FS along the antinodal direction.  The c-axis tunnelling matrix element is then responsible for ridding of the remaining (nodal) states, which leads to an insulating like behaviour.

The results of the arc model can be understood simply as a two oscillator model: one Drude at $\omega=0$, and a displaced oscillator at $\omega=\Delta$.  The dc-conductivity is the value from these two contributions at $\omega = 0$.  As the scattering rate is increased these two contributions broaden.  At first, the broadening causes a decrease in the conductivity as the Drude peak comes down. This continues until the displaced peak at the gap energy leaks into the $\omega=0$ region, leading to a gain in conductivity.  The conductivity then tends to decrease as the peaks broaden further.   This explains the weak maxima seen in both the ab- and c-axis resistivity (Figure \ref{fig:3}).  To show the two contributions explicitly we weighted the FS contribution to the conductivity by a parameter $0\le\alpha\le1$ (Figure \ref{fig:4}).  As we weaken the Drude component of the conductivity, the sample becomes increasingly more resistive.  This depletion of the remaining quasiparticles is precisely the same physics behind the resistive behaviour coming from the c-axis tunnelling matrix element.

In conclusion, we have found that the YRZ model of the pseudogap state with a linear in temperature scattering rate naturally reproduces the resistivity seen in this phase.  A simpler model, the arc model, was also shown to possess qualitatively correct features of the resistivity.  The physics in the two models which gives rise to the resistivity are very different.  In the YRZ model, it was the interplay between thermal factors and a density of states which was responsible for the behaviour of the resistivity.  By contrast, the density of states in the arc model is constant and temperature only entered through the scattering rate. However, owing to the two distinct contributions to the conductivity, the arc model still showed qualitatively correct features.  

Two elements in the YRZ calculation are vital for the agreement with experimental resistivity. First is the reconstruction of FS as a function of doping.  This reconstruction reduces the number of quasiparticles available for dc-transport as the Mott insulating phase is approached.  The second element is the tunnelling matrix element.  This matrix element effectively removes the remaining nodal quasiparticles and leaves an insulating response.  The fact that FS reconstruction is so important allows us to understand why the arc model can do well, despite lacking all of the microscopic elements of the YRZ model.  Indeed, one can think of the arc model as a phenomenological version of YRZ, where one is willing to ignore the mechanisms behind the reconstruction of the FS.  Nevertheless, the arc model can still be used to help deduce the underlying physics, and provides simple analytical results which are of great use.
 
The authors thank the Natural Sciences and Engineering Research Council of Canada (NSERC) and
the Canadian Institute for Advanced Research (CIFAR) for financial support.
\bibliography{biblio}

\end{document}